\documentclass[9pt,showpacs,letterpaper,aps,pra,twocolumn,notitlepage]{revtex4-1}
\usepackage{amsmath,amssymb,graphicx,comment}

\newcommand{\eq}[1]{\begin{align}#1\end{align}}

\newcommand{\hc}{\text{H.c.}}

\newcommand{\rr}{\mathbf{r}}
\newcommand{\E}{\mathbf{E}}
\newcommand{\B}{\mathbf{B}}
\newcommand{\D}{\mathbf{D}}
\newcommand{\del}{\mathbf{\nabla}}
\newcommand{\poly}{\text{poly}}

\begin{document}
\preprint{APS/123-QED}

\title{Why you should not use the electric field to quantize in nonlinear optics}
\author{Nicol\'as Quesada}
\affiliation{Department of Physics \& Astronomy, Macquarie University, NSW 2109, Australia}

\author{J. E. Sipe}
\affiliation{Department of Physics, University of Toronto, Toronto, ON, M5S 1A7, Canada}


\begin{abstract}
We show that using the electric field as a quantization variable in nonlinear optics leads to incorrect expressions for the squeezing parameters in spontaneous parametric down-conversion and conversion rates in frequency conversion. This observation is related to the fact that if the electric field is written as a linear combination of bosonic creation and annihilation operators one cannot satisfy Maxwell's equations in a nonlinear dielectric.
\end{abstract}

\maketitle

The quantization of the electromagnetic field in a nonlinear medium is a nontrivial task\cite{drummond14}. To
achieve this goal, one can consider the interaction between light and matter microscopically \cite{cohen98,power59,woolley71,drummond99,hillery85,hillery97} and explicitly treat the matter degrees of freedom, as has been done by a host of researchers. One can instead try to develop an effective field theory where the matter degrees of freedom are included in a phenomenological manner via susceptibilities \cite{boyd08}. The first attempts to carry out such program were developed by Born and Infeld for a linear medium in the early days of quantum mechanics \cite{born34}.
In this letter we show that any approach along these lines where the electric field is written as a linear combination of bosonic creation and annihilation operators is \emph{inconsistent} with the Maxwell equations (MEs) in the nonlinear regime, and leads to \emph{incorrect} expressions for three and higher order photon interaction terms in Hamiltonians used to study common nonlinear processes such as parametric down-conversion, frequency conversion, and four-wave
mixing.\\
\indent We start by reviewing two approaches to writing the electromagnetic energy, one using the electric field and the other the electric displacement, and then show how to provide appropriate commutation relations for the fields that will give the correct equations of motion (EOM). 
We also show that the quantization approach introduced by Hillery and Mlodinow \cite{hillery84}, formulated in terms of the vector potential and starting from a Lagrangian framework, is fully equivalent to the one introduced by Sipe \emph{et al.} \cite{sipe04}, where instead one directly writes commutation relations for the (observable) fields of the theory. Then we show that \emph{any} approach where the electric field is written as linear combination of boson operators is inconsistent with MEs and leads to expression that underestimate the strength of $n$ wave mixing processes by a factor of $n$. We conclude by reviewing other reasons why it is more convenient to work with the displacement field even in the limit of \emph{linear} optics.

To quantize a classical dynamical system from a Hamiltonian framework one must provide a Hamiltonian that is numerically equal to the energy, and a closed set of commutation relations (CR) for the operators to be associated with the relevant classical variables. The quantum EOMs follow from Heisenberg’s rule,
which for an operator $O$ yields
\eq{\label{heisenberg}
\partial_t O  = [O,H]/(i \hbar).
}
and for operators associated with the relevant classical variables these
EOMs should agree with the corresponding classical EOMs. For the
electromagnetic field said EOMs are precisely MEs. In this letter we will be interested in nonmagnetic materials that lack free charges and currents. Under these assumptions MEs are simply
\begin{subequations}
\label{MEs}
\eq{
\partial_t \mathbf{B}&=-\mathbf{\nabla} \times \mathbf{E},   \label{faraday}\\ 
\partial_t \mathbf{D} &= \mathbf{\nabla} \times (\mathbf{B}/\mu_0) ,  \label{ampere}\\
\del \cdot \D &= \del \cdot \B = 0.  \label{gauss}
}
\end{subequations}
The Eqs.
(\ref{gauss}) are not EOMs but can be thought of initial conditions on the
transversality of the fields $\D$ and $\B$; if the constraints (\ref{gauss}) are satisfied at some initial time $t_0$ , then Eqs. (\ref{faraday}-\ref{ampere}) guarantee that they are satisfied at all later times, regardless of the constitutive relation between $\D$ and $\E$. Note that the same constant transversality cannot be generally assumed for the electric field $\E$.\\
\indent To derive the Hamiltonian of the system we can start with the standard expression for the energy density  \cite{jackson07,born34}
\eq{\label{density}
\mathcal{H}=&  \int \mathbf{H} \cdot d \mathbf{B} + \mathbf{E} \cdot d \mathbf{D}.
}
The Hamiltonian is simply
\eq{\label{hamil}
H = \int  d\rr \ \mathcal{H}.
}
To carry out the integration of the Hamiltonian density one needs to specify constitutive relations. Since we assume a nonmagnetic material we trivially have $\mathbf{H} = \B/\mu_0$ , while the relation linking $\E$ to $\D$ involves the macroscopic polarization $\mathbf{P}$ of the medium
\eq{\label{const}
\mathbf{P}  = \D -\epsilon_0 \E.
}
To perform the integral (\ref{density}) one can follow one of two approaches. \\
\indent \textbf{Approach I}: Consider $\E$ the fundamental field and write the polarization in a power series in $\E$ via electric susceptibilities
\begin{subequations}
\eq{
\E \cdot d \D&= \E \cdot \left( \frac{d \D(\E)}{d\E} \right) \cdot d\E, \\
\mathbf{D}(\E) &= \mathbf{P}(\E) +\epsilon_0 \E, \\
\mathbf{P}(\E) &= \epsilon_0 \left( \chi^{(1)}\E +\chi^{(2)} \E^2+\chi^{(3)} \E^3+\ldots \right), \\
\int d\mathcal{H} &= \epsilon_0 \left(\frac{(1+\chi^{(1)})}{2}\E^2+\sum_{n\geq 2}^N \frac{n}{n+1} \chi^{(n)} \E^{n+1} \right)\nonumber \\
&+\frac{\B^2}{2\mu_0}. \label{HE}
}
\end{subequations}
In the last equation and in the rest of this letter we use $N$ to indicate the highest order nonzero susceptibility of the material.
Here and below we keep implicit the usual tensor contraction of the susceptibilities with fields, the tensor $
\chi^{(n)}$ possessing $n+1$ Cartesian components; for the energy argument to hold in this form the tensors are taken to be independent of frequency and thus to satisfy full permutation symmetry.\\
\indent \textbf{Approach II}: Consider $\D$  the fundamental field and express the polarization  in terms of $\D$ via the $\Gamma$ tensors\cite{sipe04}
\eq{
\mathbf{P}(\D) = \Gamma^{(1)} \D+\Gamma^{(2)}\D^2+\Gamma^{(3)} \D^3+\ldots
}
Equivalently one can express $\E$ directly in terms of $\D$ using inverse susceptibilities (see Chap. 1 of Drummond and Hillery \cite{drummond14})
\eq{\label{ED}
\E(\D)= \sum_{n=1}^N \eta^{(n)} \D^n.
}
The relation between the $\eta$s and the $\Gamma$s is easily read from Eq. (\ref{const}), $\epsilon_0 \eta^{(1)}=1-\Gamma^{(1)}$ and $\epsilon_0 \eta^{(n)}=-\Gamma^{(n)}, n>1$. One can also relate the usual electric susceptibilities with the inverse susceptibilities; for the two lowest order ones we have \cite{drummond14}
\begin{subequations}\label{rel}
\eq{
\eta^{(1)}_{ij} &= \epsilon_0^{-1}((1+{\chi}^{(1)})^{-1})_{ij}\\
\eta^{(2)}_{jnp} &= -\epsilon_0 \eta^{(1)}_{jk} \chi_{klm}^{(2)} \eta^{(1)}_{ln} \eta^{(1)}_{mp}.
}
\end{subequations}
Using Eq. (\ref{ED}) one arrives at the following expression for the energy density
\eq{\label{HD}
\mathcal{H}=\frac{\mathbf{B}^2}{2 \mu_0} +\sum_{n\geq 1}^N\frac{1}{n+1}\eta^{(n)} \mathbf{D}^{n+1}.
}
\\
\indent Note that for $N > 1$ the prefactors in Eq. (\ref{HE}) and Eq. (\ref{HD}), $n/(n + 1)$ and $1/(n + 1)$ respectively, are \emph{different}. We will comment on this difference shortly.
\\
\indent To complete the quantization procedure we need to provide a set of CRs that together with Heisenberg's rule Eq. (\ref{heisenberg}) will give rise to MEs (\ref{MEs}). Such commutation relations are given by \cite{sipe04}
\begin{subequations}
\eq{
[D_k(\mathbf{r}),B_l(\mathbf{r}')]&=i\hbar \epsilon_{klm} \frac{\partial}{\partial r_m} \delta(\mathbf{r}-\mathbf{r}'), \label{CR}\\
[D_k(\mathbf{r}),D_l(\mathbf{r}')]&=[B_k(\mathbf{r}),B_l(\mathbf{r}')]=0
}
\end{subequations}
where the indices $k,l,m$ denote Cartesian components, $\epsilon_{klm}$ is the Levi-Civita symbol and $\delta(\rr)$ is the Dirac distribution. It is interesting to note that these relations were originally written by Born and Infeld to quantize a linear field in 1934 \cite{born34}. Hillery and Mlodinow\cite{hillery84,drummond14} give as nonzero CR $[A_j(\mathbf{r}),\Pi_k(\mathbf{r}')]=i \hbar \delta_{jk}^{\text{tr}}(\mathbf{r}-\mathbf{r}')$ where $\Pi=-\mathbf{D}$, $\nabla \times \mathbf{A}=\mathbf{B}$, $\delta^{\text{tr}}$ is the transverse Dirac distribution and $\mathbf{A}$ is the vector potential.  But by simply taking the curl of their commutation relation with respect to $\rr$, one obtains precisely (\ref{CR}).
Using the Hamiltonian density Eq. (\ref{HD}) (the Hamiltonian is $H=\int d\rr \ \mathcal{H}$) the CR Eq. (\ref{CR}) and defining $\E$ via Eq. (\ref{ED}) it is straightforward to show that Heisenberg's rule does indeed give MEs as the EOMs of the field, and thus completes the quantization procedure sketched in the introduction.
\\
The fact that $\mathbf{D}$ and $\mathbf{B}$ are the fields that appear in the CR strongly suggest that these fields are the ones that ultimately will be written as linear combinations of bosonic creation and annihilation operators, and should be thought of as ``fundamental''.
\\
\indent Indeed, a reasonable way of settling the question of which of the two fields $\E$ and $\D$ should be thought of as fundamental
is to ask which one will consist of a linear superposition of bosonic creation and annihilation operators. If one insists on thinking that $\E$ is fundamental in a medium with a nonlinear susceptibility ($N>1$) then
\eq{
\E = &\poly_1(a_\sigma,a_\sigma^\dagger)\\
& \Rightarrow \D=\epsilon_0 \left(\E+\sum_{n=1}^N \chi^{(n)} \E^n\right) = \poly_N(a_\sigma,a_\sigma^\dagger), \nonumber
}
where $a_\sigma$ is a destruction operator for a photon with labels $\sigma$ (wavevector and  polarization in 3D, or centre frequency and waveguide mode in a 1D geometry) that satisfy the usual commutation relation $[a_\sigma,a^\dagger_{\sigma'}]=\delta_{\sigma,\sigma'}$ and $\poly_M(x)$ is a shorthand notation for a polynomial of degree $M$ in the variables $x$. 
Assuming that $\B = \poly_1(a_\sigma,a^\dagger _\sigma)$, since in a nonmagnetic material we certainly expect $\B = \mu_0 \mathbf{H}$ is a fundamental field, one has, $H=\poly_{N+1}(a_\sigma,a^\dagger _\sigma)$. Let us now calculate what Heisenberg's rule Eq. (\ref{heisenberg}) and Faraday's law Eq. (\ref{faraday}) gives us for the time evolution of $\B$ \footnote{To show how the different polynomials change of degree upon commutation one only need to remember that $[a,f(a,a^\dagger)]=\frac{\partial f}{\partial a^\dagger}, \text{ and } 
 [a^\dagger,f(a,a^\dagger)]=-\frac{\partial f}{\partial a}$}
\eq{
\partial_t \B &= [\B,H]/(i \hbar) = [\poly_1(a_\sigma,a^\dagger _\sigma),\poly_{N+1}(a_\sigma,a^\dagger _\sigma)] \nonumber \\
&= \poly_{N}(a_\sigma,a^\dagger _\sigma) \stackrel{?}{=} \del \times \E = \poly_1(a_\sigma,a^\dagger _\sigma).
}
leading to a contradiction. Thus it is not possible to have an electric field that is \emph{linear} in creation and annihilation operators and at the same time satisfy MEs in a \emph{nonlinear} medium ($N>1$). This was first noted by Hillery and Mlodinow when the expasion used for $\E$ was one obtained in terms of plane waves in a medium with no polarization. For this particular choice one would also conclude that the fields propagate at the speed of light in vacuum, which is clearly nonsense since one is trying to quantize in a dielectric \cite{drummond14}.
The derivation presented here is more general since it makes no assumptions other than the fact the photon bosonic operators satisfy their usual algebra and that a series expansion of the nonlinear polarization is appropriate.\\
It is straightforward to show that simply by choosing $\D$ as the fundamental field
\eq{
\D = \poly_1(a_\sigma,a_\sigma^\dagger) \Rightarrow \E=\sum_{n=1}^N \eta^{(n)} \D^n = \poly_N(a_\sigma,a_\sigma^\dagger)
}
one can avoid violating the MEs.\\
\indent Thus far we showed that if one insists on thinking of $\E$ as the ``fundamental'' field then one arrives at a contradiction with MEs. Let us now study in more detail what other consequences this has in the study of phase matched nonlinear optics in three-wave mixing processes. To this end, consider three modes of light, confined in the $xy$ plane by a waveguiding structure with an index profile independent
of $z$, and propagating in the $z$ direction; at each wave vector $k$  and in the absence of nonlinear interactions each mode $J$ oscillates at frequency $\omega_{Jk}$ . The spatial distribution of the fields of these modes are obtained
by solving the master equation \cite{joan11}
\eq{\label{master}
\mathbf{\nabla \times }\left( \epsilon_0 \eta^{(1)}(\mathbf{r}) \mathbf{\nabla }\times \mathbf{B}_{Jk}(%
\mathbf{r})\right) =\frac{ \omega _{Jk}^{2}}{c^{2}}\mathbf{B}_{Jk}(\mathbf{r}).
}
For propagating modes in the $z$ direction one can write $\mathbf{B}_{Jk}(\mathbf{r}) =(2\pi)^{-1/2} \mathbf{b}_{Jk}(x,y)\exp(ikz)$. 
The associated electric displacement field follows from Ampere's law (\ref{ampere}) for harmonic fields, $\D_{Jk} (\rr) =(- i \mu_0 \omega_{Jk})^{-1} \del \times \B_{Jk}(\rr)$, and is of the form $(2\pi)^{-1/2} \mathbf{d}_{Jk}(x,y)\exp(ikz)$.
The profiles $\mathbf{d}_{Jk} (x, y)$ are conveniently normalized to satisfy
\eq{\label{norm}
\int dxdy\frac{\mathbf{d}_{Jk}^{\ast }(x,y)\cdot \mathbf{d}_{Jk}(x,y)}{%
\epsilon _{0}n^{2}(x,y;\omega _{Jk})}\frac{v_{p}(x,y;\omega _{Jk})}{%
v_{g}(x,y;\omega _{Jk})}=1,
}
where $n$, $v_g$ , and $v_p$ are respectively the index of refraction, the group
velocity of mode $J$ at wave vector $k$, and its phase velocity. We have
written (\ref{norm}) in a form that is valid even if the constituent media are
dispersive \cite{landau,sipe06}, and with its adoption the field expansions
\begin{subequations}\label{expansions}
\eq{
\mathbf{B(r}) &=\sum_{J}\int dk\;\sqrt{\frac{\hbar \omega _{Jk}}{2}}a_{Jk}%
\mathbf{B}_{Jk}(\mathbf{r})+\hc,  \label{modeexpansion} \\
\mathbf{D(r}) &=\sum_{J}\int dk\;\sqrt{\frac{\hbar \omega _{Jk}}{2}}a_{Jk}%
\mathbf{D}_{Jk}(\mathbf{r})+\hc,   
}
\end{subequations}
identify that half the energy of (a freely propagating) photon is carried
by the $\D$ field and half by the $\B$ field.
In the last equation the only nonzero commutation relation for the bosonic operator is $[a_{Jk},a_{J'k'}^\dagger]=\delta_{JJ'}\delta(k-k')$.
Because the master equation (\ref{master}) is satisfied the use of the expansions (\ref{expansions}) will diagonalize the linear Hamiltonian
\eq{\label{HL}
H_L=\int d\rr \left(\frac{\B^2}{2 \mu_0}+\frac{\eta^{(1)} \D^2 }{2} \right) = \sum_{J}\int dk \ \hbar \omega_{Jk} \ a^\dagger_{Jk}a_{Jk},
}
where after the second equality we have dropped the zero point energy.
Now let us assume that three of these modes $J=A,B,C$ are phase and energy matched; that is, there are ``centre'' wavevectors in the dispersion relation of each mode that satisfy 
\eq{
\bar k_A+\bar k_B=\bar k_C \text{ and }\omega_{A \bar k_A }+\omega_{B \bar k_B}=\omega_{C\bar k_C}.
}
The nonlinear part of the Hamiltonian from Eq. (\ref{HD}) is now
\eq{\label{HNLuse}
&H_{NL}=\frac{1}{3}\int d\rr \ \eta^{(2)}\D^3\\
&=\frac{3!}{3}\int dk_{A}dk_{B}dk_{C}\sqrt{\frac{\hbar \omega
_{Ak_A}}{2}\frac{\hbar \omega _{Bk_B}}{2}\frac{\hbar \omega _{B k_C}}{2}}\nonumber \\
&a_{Ak_{A}}^{\dagger }a_{Bk_{B}}^{\dagger }a_{Ck_{C}}  \nonumber \\
&\times \int d\mathbf{r}\;\eta ^{(2)}_{ijk}(\mathbf{r})\left(
\D_{Ak_{A}}^{i}(\mathbf{r)}\right) ^{\ast }\left( \D_{Bk_{B}}^{j}(\mathbf{r}%
)\right) ^{\ast }\D_{Ck_{C}}^{k}(\mathbf{r})+\hc  \nonumber
}
where in the last equation we have used the full permutation symmetry of the $\eta^{(2)}$ tensor and hence get a combinatorial factor of $3!$. If one wishes to write the last equation in terms of the usual nonlinear susceptibilities one should make the substitution ${\eta}^{(2)}=-\epsilon_0  \chi^{(2)} {\eta_A}^{(1)} {\eta_B}^{(1)} {\eta_C}^{(1)}$. \\
\indent Let us now attempt to write the nonlinear part of the Hamiltonian by \emph{incorrectly} assuming that $\E$ is linear in the $a_{Jk}$ and that it satisfies the linear constitutive relation linking it with $\D$
\eq{\label{wE}
\tilde \E = \sum_{J}\int dk\;\sqrt{\frac{\hbar \omega _{Jk}}{2}}a_{Jk}%
\mathbf{E}_{Jk}(\mathbf{r})+\hc=  \eta^{(1)} \D,
}
where we used $\tilde \E$ to indicate that we (wrongly) kept only the linear part of the constitutive relation Eq. (\ref{ED}). Keeping this linear part will at least guarantee that the $\E$ field will move at the correct speed in the medium.\\
\indent We can now write the nonlinear Hamiltonian that would follow from this strategy by substituting $\tilde \E$ from Eq. (\ref{wE}) in Eq. (\ref{HE}), and using Eq. (\ref{rel}) we would find 
\eq{
\tilde H_{NL}=&\frac{2}{3} \epsilon_0 \int d\rr \ \chi^{(2)} \tilde \E^3 = -\frac{2}{3}\int d\rr \ \eta^{(2)} \D^3.
}
This expression has the wrong sign and is off by a factor of two with respect to the correct expression Eq. (\ref{HNLuse}). This  comes about from the difference between $n/(n+1)$ in Eq. (\ref{HE}) and $1/(n+1)$ in Eq.(\ref{HD}).
Let us try to elucidate why we get the wrong prefactor. To this end, remember once more that the correct relation between $\E$ and $\D$ is given by Eq. (\ref{ED}) and not Eq. (\ref{wE}). Since $\E$ is nonlinear in $\D$ then even the quadratic part of the Hamiltonian in terms of $\E$ in Eq. (\ref{HE}) will contain 3 photon interactions:
\eq{
\epsilon_0\frac{(1+\chi^{1})}{2}\E^2 &= \epsilon_0 \frac{(1+\chi^{(1)})}{2}  \left(\eta^{(1)}\D +\eta^{(2)}\D^2\right)^2\\
&\approx \frac{\eta^{(1)} \D^2}{2} + \eta^{(2)}\D^3.\nonumber
}
Note that the last term will, when added with $(-2/3) \eta^{(2)} \D^3$, give  $(1/3) \eta^{(2)} \D^3$, which is the correct expression Eq. (\ref{HNLuse}).
Moving to the interaction picture of Eq. (\ref{HNLuse}) with respect
to the free Hamiltonian in Eq. (\ref{HL}), we have
\eq{\label{HIcorrect}
H^I_{NL} =& \theta \int dk_{A}dk_{B}dk_{C} a_{Ak_{A}}^{\dagger }a_{Bk_{B}}^{\dagger }a_{Ck_{C}} \Phi(\Delta k L/2) e^{i\Delta t} \nonumber \\
&+\hc
}
where we assumed that the nonlinearity has a top hat profile of length $L$ in the $z$ direction and introduced
\eq{
&\Phi(\Delta k  L/2)=\int_{-L/2}^{L/2} \frac{dz}{L} e^{i \Delta k \ z} = \text{sinc}(\Delta k \ L/2),\\
\Delta k &= k_C-k_B-k_A, \\
 \Delta &= \omega_{A k_A}+\omega_{B k_B}-\omega_{C k_C},\\
\theta &= 2 L \sqrt{\frac{\hbar \omega_{A \bar k_A}}{4 \pi} \frac{\hbar \omega_{B \bar k_B}}{4 \pi} \frac{\hbar \omega_{C \bar k_C}}{4 \pi}} \times \\
&\int dx dy \eta_{ijk}^{(2)}  \left(\mathbf{d}_{Ak_{A}}^{i}(x,y)\right)^{\ast } \left( \mathbf{d}_{Bk_{B}}^{j}(x,y)\right)^{\ast } \mathbf{d}_{Ck_{C}}^{k}(x,y)\nonumber.
}
Eq. (\ref{HIcorrect}), a factor of two smaller and of opposite
sign from what would achieved by the incorrect strategy of using Eq.
(\ref{wE}), is the correct interaction Hamiltonian for spontaneous parametric down conversion and frequency conversion using three-wave mixing. If the corresponding four-wave mixing calculation is made, the correct interaction Hamiltonian is a factor of three small than what would be achieved by the incorrect strategy of using Eq. (\ref{wE}). Note that recent experiments \cite{gil17} have been able to indicate the interplay between three and four wave mixing processes and thus it becomes important to get right the different weights of the nonlinearities.

It is important to mention that it is inconvenient to think of $\E$ as
the fundamental field not only in nonlinear optics, but in linear optics as well. For instance, Eqs. (\ref{MEs}) tell us that once the initial condition of $\D$ and $\B$ being divergenceless is imposed, then the dynamics will guarantee that $\D$ and $\B$ remain divergenceless. In practice, the divergenceless of $\D$ and $\B$ can be guaranteed by expanding them in terms of modes that are divergenceless. Such a claim and such a strategy do not apply to $\E$. As well, the natural variables for the Hamiltonian density (see Eq. (\ref{density})) are $\B$ and $\D$, whereas $\mathbf{H} =\frac{\partial \mathcal{H}}{\partial \B}$ and $\mathbf{E} =\frac{\partial \mathcal{H}}{\partial \D} $ are derived fields. As pointed out by Born and Infeld \cite{born34} this fact makes $\B$ and $\D$ the natural set of fields for quantization given the fact that in nonrelativistic quantum mechanics Hamiltonians play a central role. This should be contrasted with a Lagrangian perspective where the (differential) Lagrangian density $d\mathcal{L} = -\D \cdot d\E+\mathbf{H}\cdot d\B$ has as natural variables $\E$ and $\B$ \cite{born34}.

\indent Finally, we should comment that even starting from a microscopic description of the interactions between charges and fields it is the $\D$ field that couples directly to the dipole moments of stable, neutral, non-overlapping units --molecules, atoms, quantum dots or the like. This is readily seen after applying the Power-Zienau-Woolley gauge transformation to the minimal coupling Hamiltonian \cite{power59,stenholm73,woolley71,healy77,hillery85}.

\indent In this letter we have shown very generally that it is not possible to satisfy Maxwell's equations in a nonmagnetic, nonlinear medium and at the same time write the electric field as a linear combination of creations and annihilation operators. If one insists on writing the electric field in this way one will violate Faraday's law, and as well derive incorrect expressions for the three- and four-wave mixing photon-photon interaction amplitudes that are important for parametric down-conversions, second harmonic generation, and four-wave mixing.

{\bf{Funding:}} The authors acknowledge support of the National Science and Engineering Research Council of Canada.\\

{\bf{Acknowledgment:}} NQ thanks G. Harder and J.M. Donohue for valuable discussions.

\bibliography{biblio}

\end{document}